# Potential barrier of Graphene edges

*Weiliang Wang and Zhibing Li[*]*

State Key Lab of Optoelectronic Materials and Technologies, School of Physics and Engineering, Sun Yat-sen University, Guangzhou 510275, People's Republic of China.

ABSTRACT. We calculated row resolved density of states, charge distribution and work function of graphene's zigzag and armchair edge (either clean or terminated alternatively with H, O or OH group). The zigzag edge saturated via OH group has the lowest work function of 3.76 eV, while the zigzag edge terminated via O has the highest work function of 7.74 eV. The angle-dependent potential barrier on the edge is fitted to a multi-pole model and is explained by the charge distribution.

KEYWORDS: Graphene, edge, work function.

PACS: 73.22.Pr, 73.30.+y

I. INTRODUCTION

Graphene as a unique 2-dimensional electronic system has attracted considerable attention [1-3]. Recently, well controlled edge tailoring becomes possible [4, 5]. The edge properties of graphene are important for electron transport, spin polarization, and electron/ion processes on the edge in general. There have been many studies on the peculiar edge properties of graphene [6-11]. The zigzag edge (Z-edge) and armchair edge (A-edge) being the most studied edges, have drastically different electronic properties. The Z-edge can sustain edge states and resonances that are not present in the A-edge. Another important issue of the edge is the edge potential. For use of chemical sensors, the edge potential dictates the direction of charge transfer and influences the selectivity of molecules [12]. It also affects the band lineup at the graphene–metal contact [13] and has considerable impact on device performance. Recent experiments of field emission show that the graphene is also a good field emission material [14-20]. The work function is the most important feature of the edge (surface) potential. It is known as a critical quantity in understanding field emission properties. The present paper is aimed to obtain density of states, charge distribution and the vacuum potential barrier near the edge of graphene via the *ab initio* calculation and to evaluate the work functions of armchair edge and zigzag edge of various terminating structures.

---

[*] Corresponding author email: stslzb@mail.sysu.edu.cn

Several groups had calculated the work function of single wall carbon nanotubes (SWCNT) [21-26]. Generally, the work function of SWCNT should approach that of graphene in increase of the diameter of SWCNT. However, the global edge potentials of SWCNT and graphene are very different as one has axial rotation symmetry and the other has plane symmetry. The angle dependent edge potential of a monolayer atom sheet is still lack of study. Furthermore, work functions of different edges need not be the same as the work function of an infinite graphene, since there is contribution of the local field that depends on the edge structure [27, 28]. R. Ramprasad et al. had calculated the work function of Z-edge of a narrow graphene ribbon (four chains of carbon) with different adsorbates [29]. They found that the more electropositive of the adsorbate leads to the lower work function. We will investigate the graphene ribbon with either Z-edge or A-edge. To avoid the possible error from finite width, wide ribbons with ribbons width of 3.nm (14/12 chains of carbon atoms for the Z-edge/A-edge ribbon) are used.

II. COMPUTATIONAL METHODS

The simulation was done via the Vienna *Ab inito* Simulation Package (VASP) [30]. The electron-core interactions were treated in the projector augmented wave (PAW) approximation [31]. The density functional is treated by the local density approximation (LDA) (with the Ceperly–Alder exchange correlation potential [32]). A kinetic energy cutoff (400 eV) was used. Full relaxation of magnetization was performed for spin-polarized calculations. All atoms were fully relaxed until the force on each atom is less than 0.01 eV/Å. The vacuum gaps between the graphene ribbons in both parallel and perpendicular directions ($d_g$ and $d_z$ in Fig. 1) are about 3.5 *nm*. The *k* point mesh is $1\times10\times1$.

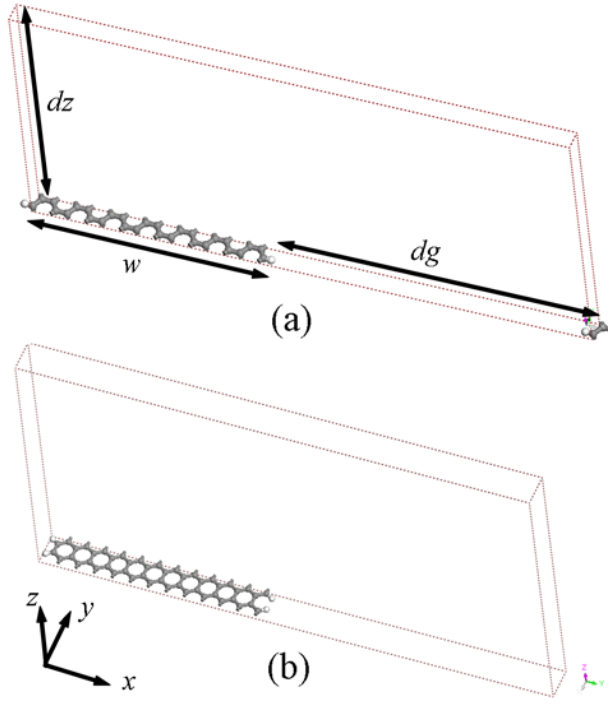

Fig. 1 Schematic illustration of the supercell of the hydrogen terminated (a) Z-edge and (b) A-edge graphene ribbon. The gray (white) balls indicate carbon (hydrogen) atoms. The width of the ribbon $w \approx$ 3. $nm$. The distance between graphene layer $d_z \approx 3.5\ nm$. The vacuum gap between edges $d_g \approx 3.5\ nm$.

Clean, H terminated, OH, and O terminated edges are investigated. Yet the OH terminated A-edge is not as stable as the O terminated one, as H and O tends to combine into H$_2$O. Therefore, we will not give the results of OH terminated A-edge. Figure 2 shows the edge structures after relaxation. Table I lists the absorption energy. The absorption energy is defined as $E_a = E_{sys} - (E_{clean} + E_{mol})$ where $E_{sys}$ is the total energy of the system with adsorbates, $E_{clean}$ is the energy of graphene with clean edge and $E_{mol}$ is the energy of adsorbates in the form of molecular (i.e. H$_2$, O$_2$ or H$_2$O).

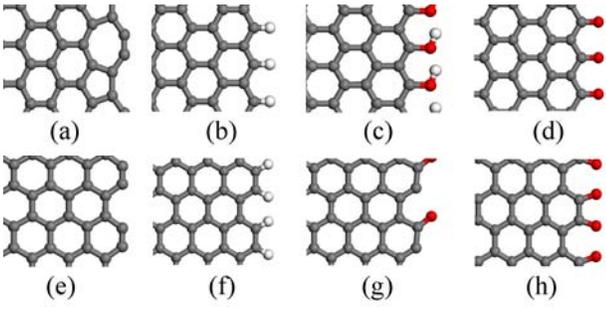

(a) (b) (c) (d)
(e) (f) (g) (h)

Fig. 2 (Color online) Edge structures. (a) reconstructed clean Z-edge [33]; (b) H terminated Z-edge; (c) OH terminated Z-edge; (d) O terminated Z-edge; (e) clean A-edge; (f) H terminated A-edge; (g) half O terminated A-edge; (h) fully O terminated A-edge.

Table I. Absorption energy. The absorption energy is defined as $E_a = E_{sys} - (E_{clean} + E_{mol})$ where $E_{sys}$ is the total energy of the system with adsorbates, $E_{clean}$ is the energy of graphene with clean edge and $E_{mol}$ is the energy of adsorbates in the form of molecular.

| Structure (label as in Fig. 2) | Total energy (eV) | Absorption energy (eV) |
|---|---|---|
| $O_2$ | -10.49 | \ |
| $H_2$ | -6.62 | \ |
| $H_2O$ | -14.84 | \ |
| clean unreconstructed Z-edge (28C) | -276.13 | \ |
| Z-edge (28C+2H) (b) | -288.97 | 6.22 |
| Z-edge (28C+2OH) (c) | -304.66 | 11.42 |
| Z-edge (28C+2O) (d) | -294.90 | 8.28 |
| clean A-edge (48C) (e) | -475.36 | \ |
| A-edge (48C+4H) (f) | -497.77 | 9.17 |
| A-edge (48C+4OH) | -519.66 | unstable |
| A-edge (48C+2O) (g) | -491.05 | 5.20 |
| A-edge (48C+4O) (h) | -506.52 | 10.17 |

The usual definition for the work function [27] is not applicable for graphene's edge, as the crystal face scale is the same as the atomic scale at the edge. Therefore, we define the work function as $WF_m = \phi_m - E_f$, where $E_f$ is the Fermi level and $\phi_m$ the maximum of vacuum potential energy in the path of electron moving out of graphene to vacuum (i.e., the barrier height in field emission).

III. RESULTS AND DISCUSSION

Figure 3 is a typical potential distribution around the graphene. The blue region is the graphene ribbon. Repeat replicas in both *x* and *z* direction should be understood. The potential in vacuum near the center of the graphene ribbon is different from the potential in vacuum in the vicinity of the edge. Therefore the work function of the edge is different from that of the center. The present paper focuses on the edges.

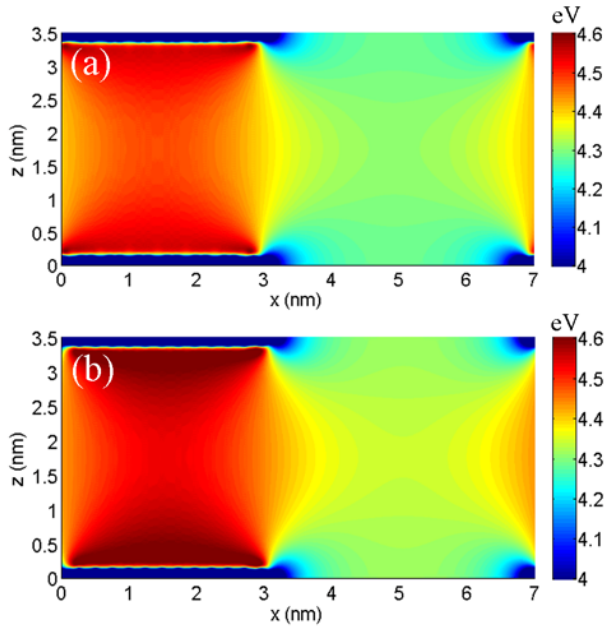

Fig. 3 (Color online) Potentials around (a) Z-edge and (b) A-edge graphene ribbons terminated by H. The blue region is the graphene ribbon.

We calculated the density of states (DOS) projected onto each atom. Figure 4 and 5 are the DOS on each row of atoms. The amount of edge states in the Z-edge with adsorbates is almost the same as the clean Z-edge. It indicates that the edge states in the Z-edge are not originated from the dangling bonds.

The edge states are also found in the clean and O terminated A-edge. That are originated from the dangling bonds as the edge state is absent in the H terminated A-edge which has no dangling bond. Our results confirm the tight binding result of Refs. [6, 8]. There is little DOS near the Fermi level on H atoms. That should have significant consequence on field emission: edge states above the Fermi level will stop the potential barrier being lowered by applied field[34] which is unfavorable for field emission; on the other hand, edge states below the Fermi level can contribute to field emission.

The electron distribution in the edge is presented in Fig. 6. The red regions in Fig. 6 (c, d, g, and h) show that the electrons are strongly concentrated at the O atoms. Comparing the electron densities in the C-O bonds and the C-C bonds we see that certain amount of electrons has transferred from C atoms to O atoms. The red arrows indicate the positions of H atoms. The electrons have shifted from H atoms to the C-H bonds in the H terminated edge (Fig. 6 b and f). In the OH edge (Fig. 6 c), the electrons have shifted from H atoms to the O-H bonds. The inhomogeneous charge distribution by the edge alternates the edge potential by producing local field at the edge. Figure 7 shows the potential energies in vacuum in the vicinity of the eight edges as defined in Fig. 2. They are plotted along the forward path that is the line perpendicular to the edge and parallel to the graphene plane, with the origin at the outer most O or C atom. It is found that the H adsorbate reduces the work function by forming a line of dipoles whose direction is pointing outwards from the graphene, while the O adsorbate raises the work function by forming an opposite dipole-line. The dipole picture is consistent with the displacement of electrons with respect to the H atomic core and to the C atomic core. An obvious potential peak near the last atom is created by the O adsorbate.

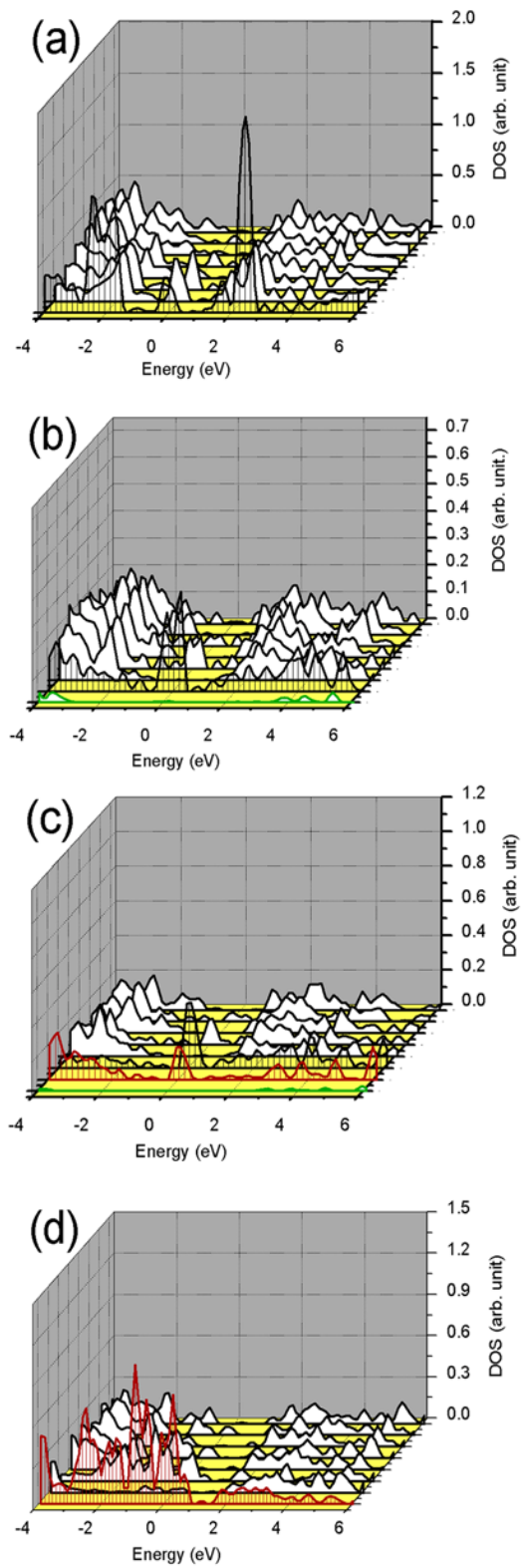

Fig. 4 (Color online) DOS in each row of atoms of (a) reconstructed clean Z-edge; (b) H terminated Z-edge; (c) OH terminated Z-edge; (d) O terminated Z-edge. The black lines are for C atoms, the red lines for O atoms, the green lines for H atoms. The energy is shifted to let the Fermi level equal to zero.

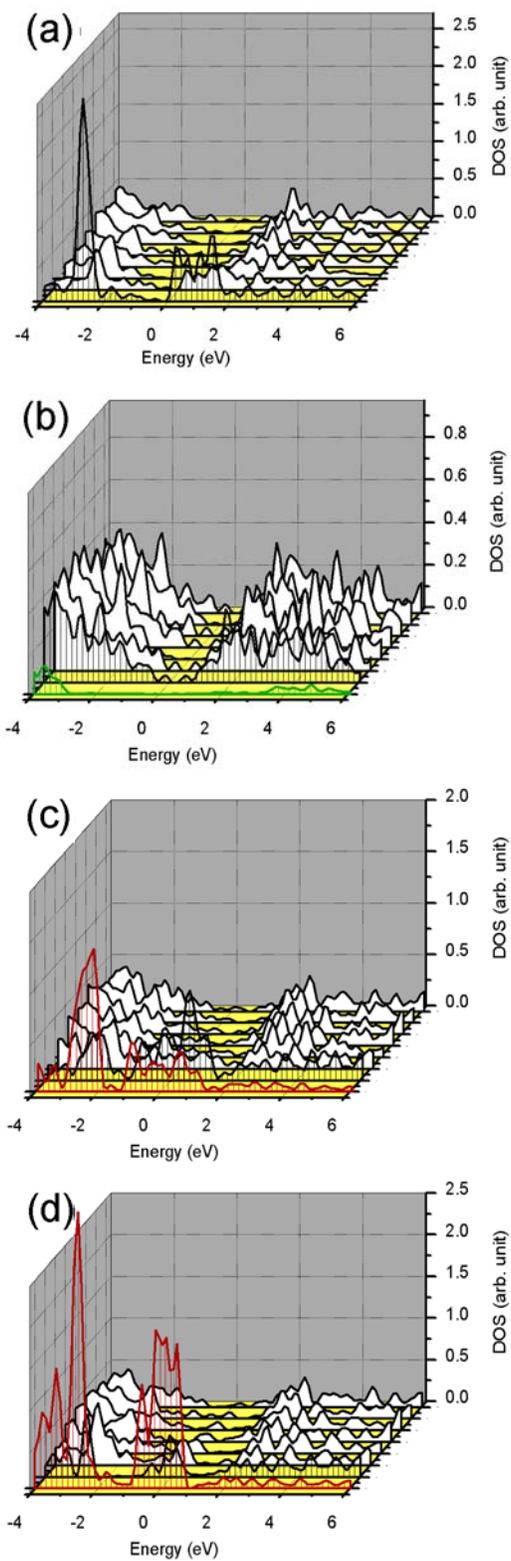

Fig. 5 (Color online) DOS in each row of atoms of (a) clean A-edge; (b) H terminated A-edge; (c) half O terminated A-edge; (d) fully O terminated A-edge. The black lines are for C atoms, the red lines for O atoms, the green lines for H atoms. The energy is shifted to let the Fermi level equal to zero.

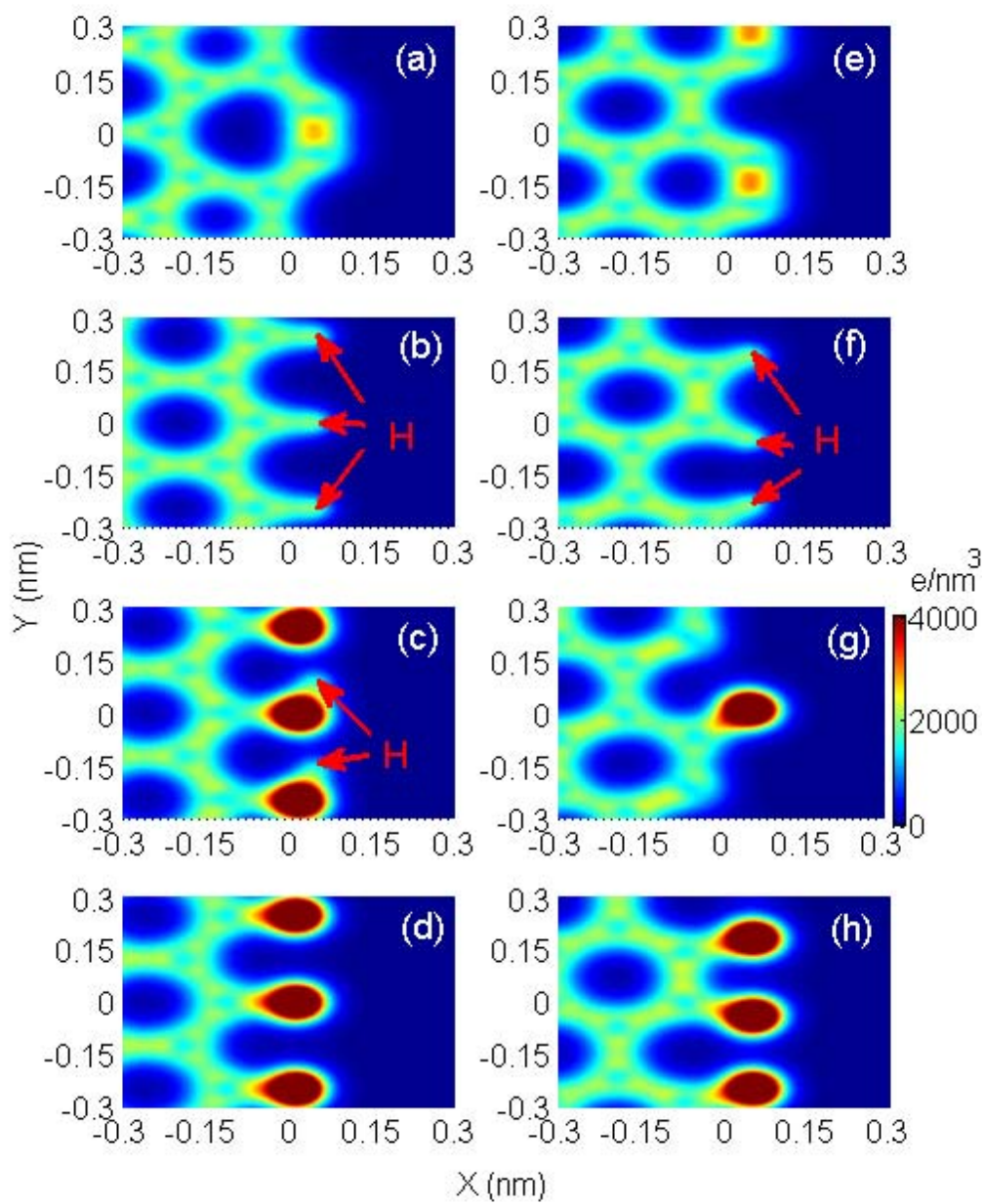

Fig. 6 (Color online) Charge distribution. (a) reconstructed clean Z-edge; (b) H terminated Z-edge; (c) OH terminated Z-edge; (d) O terminated Z-edge; (e) clean A-edge; (f) H terminated A-edge; (g) half O terminated A-edge; (h) fully O terminated A-edge. The red arrows indicate the positions of H adsorbate.

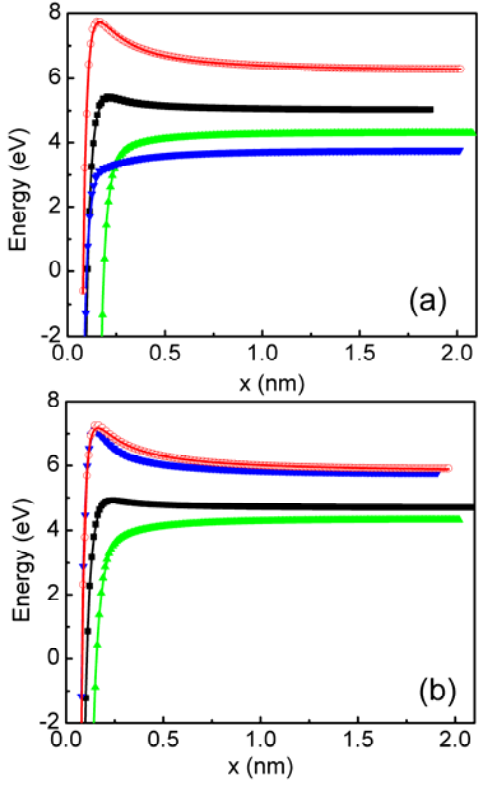

Fig. 7 (Color online) (a) Potential energies in vacuum in the vicinity of clean (black squares), H terminated (green triangles), OH terminated (blue inverted triangles) and O terminated (red circles) Z-edge of graphene. (b) Potential energies in vacuum in the vicinity of clean (black squares), H terminated (green triangles), half O terminated (blue inverted triangles) and fully O terminated (red circles) A-edge of graphene. The energies are shifted to let the Fermi levels be zero. It is plotted along a line perpendicular to the edge and parallel to the graphene plane with the origin at the outer most O or C atom.

We find that the potential energy ($u$) on the plane perpendicular to the graphene and the edge can be well fitted by a multi-dipole model,

$$u(\theta, r) = c - \frac{p}{r^2}\cos\theta - \frac{d_x(3\cos^2\theta - 1) + d_z(3\sin^2\theta - 1)}{2r^3} - \frac{f}{r}\cos\theta - \frac{g}{r} \quad (1)$$

where the first term is a constant that originates from the chemical bonding, the second term is the potential of a local dipole near the edge atom under consideration, the third term is from a local quadrupole (at the same position of the dipole), the fourth term is the contribution of a dipole-line along the edge, the fifth term is from a point charge. Here $r$ and $\theta$ are the polar coordinates with the origin at the position of the dipole, with polar axis along x axis. The fitted potentials for various edge arrangements are shown in Fig. 7 and 8 as solid curves. The five parameters $c$, $p$, $2d_x-d_z$, $f+g$, and the position of the dipole ($x_0$, not appearing in (1)), are first obtained via fitting the potential energy along the forward path (Fig. 7). There are still two adjustable parameters (say $d_x$ and $g$) which can be used to fit the potential energy in various polar angles. In Fig. 8, the potential energies versus $\theta$ at three values of $r$ (0.3, 0.6, and 0.9 nm) for each edge structure are presented. The scatters are DFT results and the solid curves are calculated with Eq. (1). The multi-dipole model reproduces the potential quite well when $|\theta|$ is smaller than 30°. The potential energy in the region $|\theta|<30°$ is what the present paper most concerned about, as most of electrons emit along the forward path. The difference between the DFT and the model is less than 0.06 eV when $|\theta|$ is smaller than 30°. Table II lists the fitting parameters and the work functions.

The potential energies of the H and OH terminated edges (which are more electropositive) increase with the angle, while the O terminated (which are more electronegative) have inversed angle-dependence and that of the clean edges are less angle dependent. The angle dependent edge potential can be mainly explained by the dipole-line (parameter $f$ in Table II). The strengths of dipole-line $f$ obtained by fitting the angle-dependent potential for various edges further confirm the dipole picture that we have seen in Fig. 6. The dipole-lines with outwards direction are responsible to the concave angle dependence of the potential. The O terminated edges and the clean edges form dipole-lines with inwards direction; therefore their potentials depend on the angle in the reverse trend.

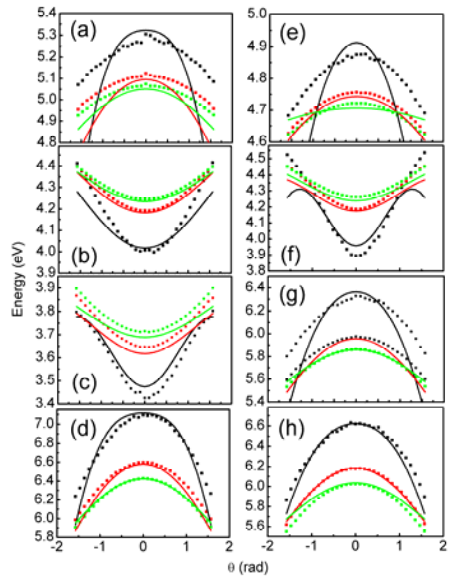

Fig. 8 (Color online) Potential energies in vacuum in the vicinity of (a) reconstructed clean Z-edge; (b) H terminated Z-edge; (c) OH terminated Z-edge; (d) O terminated Z-edge; (e) clean A-edge; (f) H terminated A-edge; (g) half O terminated A-edge; (h) fully O terminated A-edge. The solid lines are produced by Eq. (1) with $r = 0.3$ *nm* (black), 0.6 *nm* (red) and 0.9 *nm* (green).

Table II. Fitting parameters in Eq. (1) and the work functions. The position $x_0$ is the distance of the dipole from the last C or O. The energy zero is set to the Fermi level. The work function is defined as the maximum of the potential barrier along the forward path.

| Edge structure (labled as in Fig. 2) | $c$ (eV) | $p$ (eV·$nm^2$) | $dx$ (eV·$nm^3$) | $dz$ (eV·$nm^3$) | $f$ (eV·$nm$) | $g$ (eV·$nm$) | $x_0$ ($nm$) | WFm (eV) |
|---|---|---|---|---|---|---|---|---|
| Reconstructed Z-edge (a) | 5.09 | -0.1093 | 0.0166 | 0.0056 | -0.0690 | 0.2113 | 0.0060 | 5.43 |
| Z-edge+H (b) | 4.36 | -0.0209 | 0.0073 | 0.0071 | 0.1462 | -0.0148 | 0.0988 | 4.29 |
| Z-edge+OH (c) | 3.90 | -0.0498 | 0.0046 | -0.0010 | 0.1640 | 0.0723 | 0.0106 | 3.76 |
| Z-edge+O (d) | 6.20 | -0.0699 | 0.0020 | -0.0129 | -0.3713 | 0.2339 | 0.0057 | 7.74 |
| Clean A-edge (e) | 4.75 | -0.1066 | 0.0254 | 0.0211 | 0.0773 | 0.0657 | 0.0029 | 4.91 |
| A-edge+H (f) | 4.47 | -0.1071 | 0.0251 | 0.0129 | 0.2438 | 0.0595 | 0.0254 | 4.32 |
| A-edge+O (g) | 5.77 | -0.0782 | 0.0078 | 0.0004 | -0.1826 | 0.1839 | 0.0106 | 7.11 |
| A-edge+2O (h) | 5.76 | -0.0207 | 0.0001 | -0.0067 | -0.3132 | 0.0846 | 0.0207 | 7.26 |

## IV. CONCLUSION

We obtained the density of states projected on atoms at graphene's zigzag and armchair edges (either clean or terminated alternatively with H, O, and OH group). The edge states are found for all the cases except the A-edge terminated by hydrogen. There is little DOS near Fermi level on H atoms. The lowest work function of 3.76 eV is found in the zigzag edge terminated with OH. The H adsorbate reduces the work function by forming a line of dipoles whose direction is pointing from graphene to vacuum, while the O adsorbate raises the work function by forming an opposite dipole-line. The work function of the O-terminated Z-edge is as high as 7.74 eV. That agrees with Ramprasad and Allmen's conclusion that the more electropositive of adsorbates leads to lower work function [29]. However, the work function values of wider ribbons (12 to 14 carbon chains) are significantly different from that of the narrow ribbon (the zigzag graphene ribbons with 4 carbon chains) of Ref. [29].

The angle-dependent edge potential is calculated via DFT and described via the multi-dipole model. It is found that the potential of various edges can be well described by the model that contains a local charge, dipole, and quadrupole, as well as a dipole-line along the edge. Remarkably, the potential energies of the H and OH terminated edges (which are more electropositive) increase with the angle magnitude, while the O terminated (which are more electronegative) and the clean edges have inversed angle-dependence. The large variation of the edge potential for various edge structures suggests that the graphene could be a good molecular sensor.

ACKNOWLEDGMENT. The project is supported by the National Basic Research Programme of China (2007CB935500), the China Postdoctoral Science Foundation (20100470974) and the high-performance grid computing platform of Sun Yat-sen University.